\documentclass[twocolumn,amsmath,amssymb,showpacs,superscriptaddress]{revtex4-2}
\usepackage{graphicx, hyperref}
\usepackage{dcolumn}
\usepackage{bm}
\usepackage{amsmath}
\usepackage{amsfonts}
\usepackage{amssymb}
\allowdisplaybreaks
\makeatletter

\newcommand{\pprime}{{\prime\prime}}

\begin{document}
\title{Kinetic theory of dilute granular gases having an inverse power law potential}
\author{Satoshi Takada\footnote{takada@go.tuat.ac.jp}}
\affiliation{Department of Mechanical Systems Engineering and Institute of Engineering,
Tokyo University of Agriculture and Technology, 
2-24-16 Naka-cho, Koganei, Tokyo 184-8588, Japan}
\date{\today}

\begin{abstract}
The kinetic theory of dilute granular gases having an inverse power law repulsive potential is studied.
We derive the time evolution of the temperature and the transport coefficients from the Boltzmann equation.
We also investigate the linear stability analysis of the hydrodynamics, and study the softness dependence of the thresholds for the shear and heat modes against the restitution coefficient.
\end{abstract}
\maketitle

\section{introduction}
To understand the behavior of granular materials is important in both science and industry.
One of the most successful treatments of granular materials is the kinetic theory of dilute inelastic gases.
Among them, the kinetic theory of inelastic hard-core gases is well studied \cite{Savage1981,Jenkins1983,Lun1984,Jenkins1985,Brey1998,Garzo1999,Huthmann2000,Brilliantov}.
When the homogeneous cooling state is realized, the decrease of the temperature due to inelastic collisions is proportional to the inverse square of the time in the later stage, which is known as the Haff's law \cite{Haff1983}.
Brey {\it et al.}\ \cite{Brey1998} have derived the transport coefficients for dilute inelastic hard-core gases and obtained the hydrodynamics, and later, Garz\'{o} and Dufty \cite{Garzo1999} have derived those for moderately dense gases.
There are some studies considering hard-core gases with a velocity dependent \cite{Poeschel2003,Takada2017, Takada2022} or a stochastic \cite{Serero2015} restitution coefficient.
For the former system, the time evolution of the temperature was found to deviate from the Haff's law \cite{Poeschel2003} and the transport coefficients were also derived \cite{Takada2017, Takada2022}.

One of the extensions of the kinetic theory to other potentials is that of repulsive hard-core with the square-well.
The transport coefficients for elastic system were derived almost seventy years ago \cite{Holleran1951,Longuet-Higgins1958,Davis1961,Davis1965,Altenberger1975,Hirschfelder}, which is known to reproduce the temperature dependence of the shear viscosity of molecules such as nitrogen \cite{Hirschfelder}.
Recently, this theory was extended to inelastic systems \cite{Takada2015}.
In Ref.\ \cite{Takada2015}, the kinetic theory was found to be invalid for the low temperature regime even if the inelasticity is sufficiently small.
In the low temperature regime, there appear some pairs of particles trapped into the attractive well, and the assumption of two body collisions may be violated.

Another extension is the kinetic theory for soft core particles.
Especially, there are a lot of papers studying elastic and inelastic Maxwell molecules \cite{Bobylev2000,Ben-Naim2000,Ernst2002,Santos2003,Garzo2005,Khalil2014,Chapman,Hirschfelder}, where the intermolecular potential is given by inverse fourth power potential.
For Maxwell molecules, the collision frequency is known to be independent of the temperature, and in this case, the treatment of the Boltzmann equation becomes simpler.
Santos \cite{Santos2003} has derived the transport coefficients for inelastic system and hydrodynamics up to Navier-Stokes order.
Later, Khalil {\it et al.}\ \cite{Khalil2014} have considered the generalized Maxwell model, where the temperature dependence of the collision frequency is generalized, and derived hydrodynamics of this model up to Burnett order.
Alternative kinetic descriptions based on Grad's moment method have also been developed for dilute and moderately dense granular gases, providing complementary expressions for transport coefficients and hydrodynamic equations \cite{Risso02, Kremer11, Garzo13, Kremer24, Barbera25, Gupta2020}.
In particular, Gupta \cite{Gupta2020} developed moment theories for dilute granular gases of inelastic Maxwell molecules and investigated their hydrodynamic stability. 
Although the quantities considered there partly overlap with those studied here, the underlying kinetic model and theoretical approach are different. 
Gupta's analysis is based on the inelastic Maxwell model and Grad's moment method, whereas the present study explicitly incorporates the scattering process associated with the inverse-power-law potential into the Boltzmann collision operator and employs the Chapman-Enskog expansion.

In this paper, we try to construct the kinetic theory of dilute granular gases having an inverse power law repulsive potential and investigate the effect of softness on the potential.
For example, the intermolecular potentials of methane and hydrogen are approximately described by inverse sixth and inverse twelfth potentials, respectively, and the kinetic theory well reproduces the temperature dependence of the transport coefficients \cite{Chapman}.
We also note that because the scattering process for this potential is analytically or numerically solvable depending on the exponent of the potential \cite{Whittaker,GoldsteinMechanics}, we can expect to obtain simple expressions of the transport coefficients.
We also try to investigate the linear stability of the hydrodynamic equations and derive the restitution coefficient dependence of the thresholds of the wave number.

While extensive studies have been conducted for the
inelastic hard-sphere model and the inelastic Maxwell model, much less is known about granular gases interacting through general inverse-power-law potentials.
The present study provides a unified description covering the entire range $4 \le \alpha < \infty$ and clarifies how the softness of the interaction potential affects the cooling dynamics, transport coefficients, and hydrodynamic stability.
 
The organization of this paper is as follows:
In the next Section, we explain our model and study the homogeneous cooling state and derive the transport coefficients from the Boltzmann equation.
We also derive the hydrodynamics up to Navier-Stokes order and study the linear stability analysis.
In Sects.\ \ref{sec:discussion} and \ref{sec:summary}, we discuss and summarize our results.
In Appendix \ref{sec:scattering}, we briefly explain the scattering process for particles having an inverse power law potential.
In Appendix \ref{sec:BasicIntegral}, we define the Basic integral to make our calculations simpler.

\section{Boltzmann equation}\label{sec:Boltzmann}
In this section, we study the homogeneous cooling state and derive the transport coefficients from the Boltzmann equation.
We also investigate the linear stability of this homogeneous state.
Let us consider the particles whose mass is $m$ and whose interaction is given by the inverse $\alpha$ law potential as
\begin{align}
    U(r)= \varepsilon\left(\frac{d}{r}\right)^\alpha,
\end{align}
where $\varepsilon$ represent the magnitude of the repulsive potential and $d$ is the effective diameter of the particle.
For $\alpha\to \infty$, this potential reduces to the hard-core potential.
Let us consider the distribution function $f(\bm{r},\bm{v},t)$ satisfying the Boltzmann equation
\begin{align}
    \left(\frac{\partial}{\partial t}
    +\bm{v}\cdot \bm\nabla\right)f(\bm{r},\bm{v},t)
    =I(f,f)\label{eq:Boltzmann_eq}
\end{align}
with the collision integral
\begin{align}
    I(f,f)
    &=\int \mathrm{d}\bm{v}_2 \int d\hat{\bm{k}} \sigma(\chi,\bm{v}_{12}) v_{12}\nonumber\\
    &\hspace{1em}\times\left[\chi_e f(\bm{r},\bm{v}_1^\pprime,t)f(\bm{r},\bm{v}_2^\pprime,t) - f(\bm{r},\bm{v}_1,t)f(\bm{r},\bm{v}_2,t)\right],
\end{align}
where the relationship between the pre-collisional velocities $\bm{v}_1^\pprime, \bm{v}_2^\pprime$ and the post-collisional velocities $\bm{v}_1, \bm{v}_2$ is given by
\begin{align}
\begin{cases}
    \displaystyle \bm{v}_1 
    = \bm{v}_1^\pprime -\frac{1+e}{2}(\bm{v}_{12}^\pprime \cdot \hat{\bm{k}})\hat{\bm{k}},\\
    \displaystyle \bm{v}_2 
    = \bm{v}_2^\pprime +\frac{1+e}{2}(\bm{v}_{12}^\pprime \cdot \hat{\bm{k}})\hat{\bm{k}},
\end{cases}\label{eq:pre_post}
\end{align}
with $\bm{v}_{12}=\bm{v}_1-\bm{v}_2$, $\chi$ is the deflection angle, $\sigma(\chi,\bm{v}_{12})$ is the scattering cross section, $\chi_e$ is related to the Jacobian of the transformation between $(\bm{v}_1^\pprime, \bm{v}_2^\pprime)$ and $(\bm{v}_1, \bm{v}_2)$, and $e$ is the restitution coefficient independent of the relative velocity.
Here, $\hat{\bm{k}}$ denotes the unit vector along the line connecting the two particles at their closest approach during a binary collision.
It is noted that the relationship between pre-collisional and post-collisional velocities is more complicated for real situations, for example, acceleration or deceleration should be considered as discussed in Ref.\ \cite{Takada2015}.
This means that the relationship (\ref{eq:pre_post}) is a toy model.
In this paper, however, we focus on this collision rule even in this system.
This is because we can expect to obtain various features different from that of hard-core gases, and this can be regarded as a simple model of the approximation as discussed later.
Indeed, the inelastic Maxwell models ($\alpha=4$) are well studied. 
The impact parameter and the velocity dependence of the scattering angle are discussed in Appendix \ref{sec:scattering}.
We also note that the present $\alpha=4$ case differs from the conventional inelastic Maxwell model often used in kinetic theory~\cite{Santos2003, Khalil2014, Gupta2020}.
In the latter model, the collision frequency is assumed to be independent of the relative velocity, whereas the present collision kernel is directly derived from the inverse-fourth power interaction potential.
Therefore, the temperature dependence of the cooling rate is generally different.

\subsection{Homogeneous cooling state}
\begin{figure}[htbp]
    \begin{center}
	\includegraphics[width=85mm]{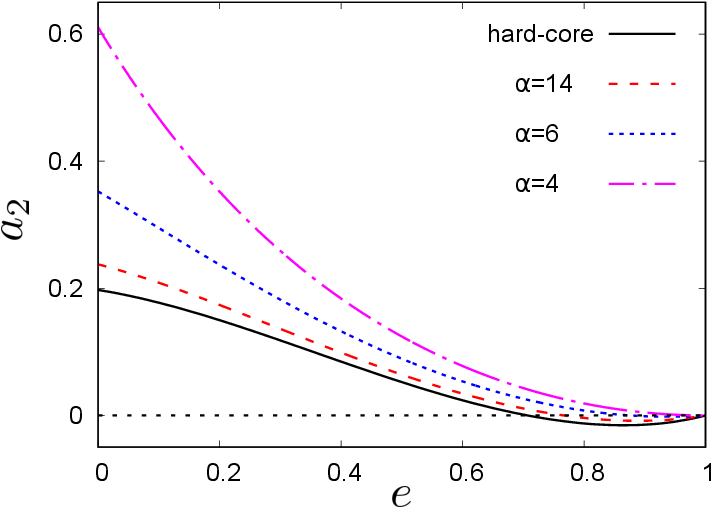}
    \end{center}
    \caption{Plot of $a_2$ against the restitution coefficient $e$ given by Eq.\ (\ref{eq:a2}) for $\alpha=14$ (red dashed line), $6$ (blue dotted line), and $4$ (pink dot-dashed line).
    The solid line shows $a_2$ for inelastic hard-core gases.}
    \label{fig:a2}
\end{figure}

We expand the velocity distribution function in terms of the Sonine polynomials as
\begin{equation}
    f^{(0)}(\bm{v},t)=f_{\rm M}(\bm{V})\left[1+a_2 S_2\left(\frac{mV^2}{2T}\right)\right],\label{eq:f0}
\end{equation}
where $V=\bm{v}-\bar{\bm{v}}$ is the local fluctuation of the velocity with the average velocity $\bar{\bm{v}}$, $f_{\rm M}(V)=n(m/2\pi T)^{3/2}\exp(-mV^2/2T)$ is the Maxwell distribution function and $S_2(x)\equiv x^2/2-5x/2+15/8$ is the Sonine polynomials \cite{Brilliantov, Brey1998, Garzo1999, Takada2015}.
Let us define the $n$-th moment $\mu_n$ as
\begin{align}
    \mu_n 
    &=-\frac{1}{2}\int \mathrm{d}\bm{c}_1 \int \mathrm{d}\bm{c}_2 \int \mathrm{d}\hat{\bm{k}}\tilde{\sigma}(\chi,c_{12}) c_{12}\nonumber\\
    &\hspace{1em}\times\tilde{f}^{(0)} (\bm{c}_1) \tilde{f}^{(0)} (\bm{c}_2) \Delta[c_1^n+c_2^n],\label{eq:mu_n}
\end{align}
where $\bm{c}_i\equiv \bm{v}_i/v_{\rm T}$ with the thermal velocity $v_{\rm T}=\sqrt{2T/m}$, $\chi$ is the deflection angle, $\tilde{f}(\bm{c}_i)=(v_{\rm T}(t)^3/n)f(\bm{v}_i)$ with the number density $n$, $\Delta[\psi(\bm{c})]\equiv \psi(\bm{c}^\prime)-\psi(\bm{c})$ is the difference between pre- and post-collisional velocities.
Here and in what follows, the angular brackets denote the average with respect to the normalized velocity distribution, $\langle A\rangle\equiv \int \mathrm{d}\bm{c}A(\bm{c})\widetilde{f}^{(0)}(\bm{c})$.
With the aid of the Basic integral shown in Appendix \ref{sec:BasicIntegral}, we can obtain the second and fourth moments as
\begin{subequations}\label{eq:mu2_mu4}
\begin{align}
    \mu_2 &= \sqrt{2\pi} \left(\frac{\varepsilon}{T}\right)^{2/\alpha}\Gamma\left(3-\frac{2}{\alpha}\right)(S_1+a_2S_2),\label{eq:mu2}\\
    \mu_4 &= \sqrt{2\pi} \left(\frac{\varepsilon}{T}\right)^{2/\alpha}\Gamma\left(3-\frac{2}{\alpha}\right)(T_1+a_2T_2),\label{eq:mu4}
\end{align}
\end{subequations}
where the explicit forms of $S_1$, $S_2$, $T_1$ and $T_2$ are given by Eqs.\ (\ref{eq:S1_S2}) and (\ref{eq:T1_T2}), respectively, in terms of 
\begin{equation}
    A_{\alpha,j} \equiv \int_0^\infty \mathrm{d}\tilde{b}\hspace{0.2em} \tilde{b} \cos^j\theta_\alpha    
\end{equation}
(the detailed derivation is shown in Appendix \ref{sec:BasicIntegral}).
Here, the quantities $\mu_2$ and $\mu_4$ denote scalar collisional moments associated with the second and fourth velocity moments, respectively.
They correspond to the trace of the second-order moment and the double trace of the fourth-order moment.
These two moments are related to each other as $(4/3)\mu_2 \langle c^4\rangle=\mu_4$ with $\langle c^4\rangle=(15/4)(1+a_2)$ \cite{Brilliantov}.
From this relationship with Eqs.\ (\ref{eq:mu2_mu4}), we can evaluate the coefficient $a_2$ as
\begin{align}
    a_2 &= \frac{T_1-5S_1}{5S_1+5S_2-T_2},\label{eq:a2}
\end{align}
where the explicit form is not shown here because of its complicated form.
It is noted that because the quantities $A_{\alpha,2}$ and $A_{\alpha,4}$ are both pure numbers, we can numerically evaluate them as calculated in the previous studies \cite{Chapman1922,Chapman,Resibois}.
It should be noted that for $\alpha\to \infty$, $A_{\alpha,2}$ and $A_{\alpha,4}$ reduce to $1/4$ and $1/6$, respectively, and then $a_2$ reduces to that for hard-core gases $
a_2^{(\rm HC)}=16(1-e)(1-2e^2)/[81-17+30e^2(1-e)]$ \cite{Brilliantov}.
Figure \ref{fig:a2} shows the restitution coefficient dependence of $a_2$.
The homogeneous cooling state of this model deviates from the Maxwell distribution function more than that of inelastic hard-core model for small $e$, and the deviation becomes smaller and smaller for large $\alpha$.
We also note that $a_2$ for $\alpha=6$ and $14$ becomes negative for small inelasticity.
This behavior is qualitatively similar to the case for hard-core gases because $a_2^{(\rm HC)}$ is negative for $e>1/\sqrt{2}$.   
On the other hand, $a_2$ for $\alpha=4$ is always positive for $e<1$.

\begin{figure}[htbp]
    \begin{center}
	\includegraphics[width=85mm]{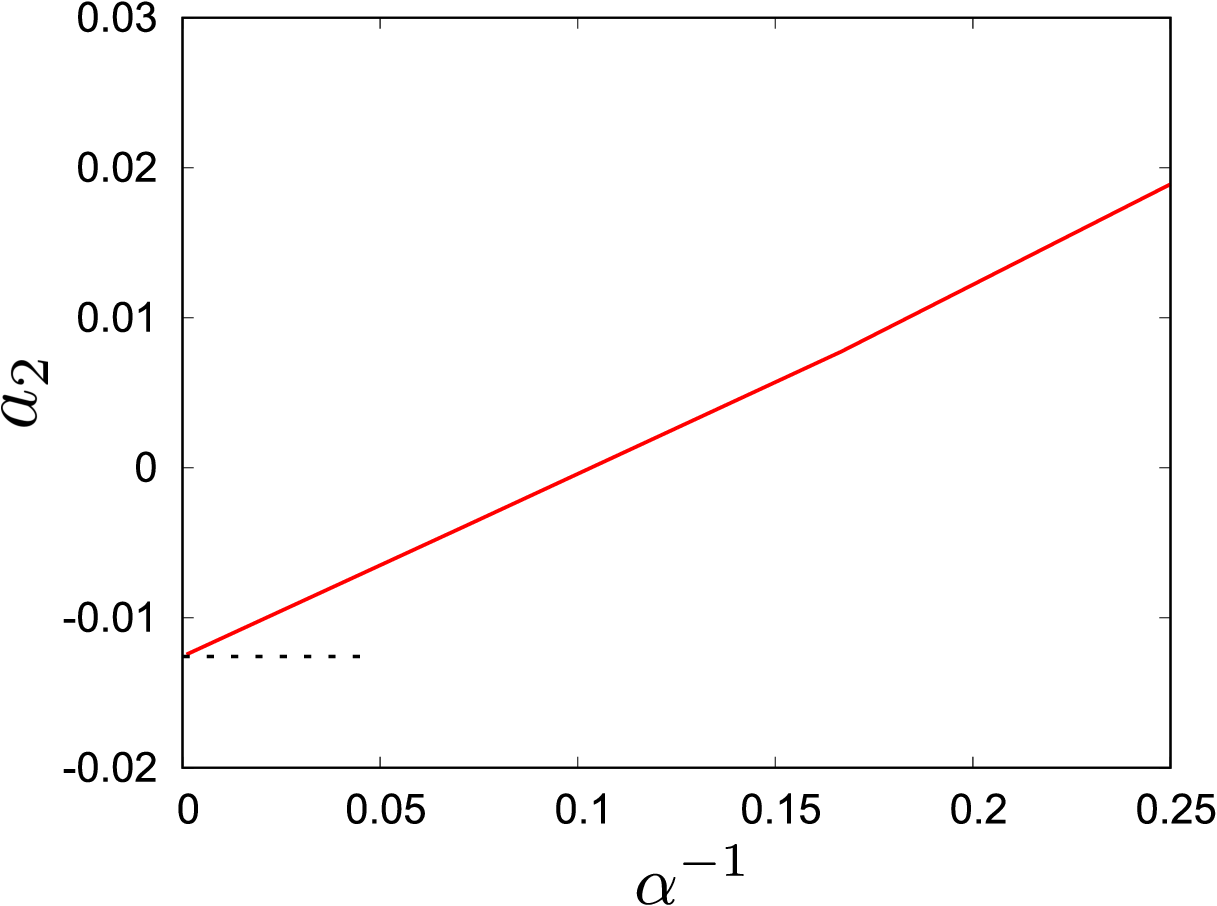}
    \end{center}
    \caption{Plot of $a_2$ against the inverse of $\alpha$ for $e=0.8$.
    The dashed line indicates $a_2^{(\rm HC)}$ for inelastic hard-core gases.}
    \label{fig:a2_alpha}
\end{figure}

Figure \ref{fig:a2_alpha} shows the dependence of the fourth cumulant $a_2$ on $1/\alpha$ for a fixed restitution coefficient $e=0.8$. 
As the interaction exponent $\alpha$ increases, the solution continuously approaches the inelastic hard-sphere (IHS) limit, while $\alpha=4$ corresponds to the inverse-fourth-power interaction, which shares the characteristic temperature scaling of Maxwell molecules.
This result clearly demonstrates that the present inverse-power-law model provides a continuous interpolation between the inverse-fourth-power and IHS limits, allowing us to systematically investigate the influence of interaction softness on the velocity distribution.

Let us derive the time evolution of the temperature for homogeneous cooling state.
From the second moment of the Boltzmann equation, the time evolution is given by
\begin{equation}
    \frac{\mathrm{d}T}{\mathrm{d}t}=-\zeta T,\label{eq:dTdt}
\end{equation}
with $\zeta=(2/3)nd^2\sqrt{2T/m}\mu_2$.
For $\alpha\neq4$, Eq.\ (\ref{eq:dTdt}) can be solved analytically with the initial condition $T=T_0$ at $t=0$ as
\begin{equation}
    T(t)
    =\frac{T_0}{\displaystyle \left(1+\frac{t}{\tau_\alpha}\right)^{2\alpha/(\alpha-4)}},\label{eq:T_t}
\end{equation}
where the characteristic time $\tau_\alpha$ is defined by
\begin{align}
    \tau_\alpha^{-1} &= \frac{4(\alpha-4)}{3\alpha}nd^2\sqrt{\frac{\pi T_0}{m}}\left(\frac{\varepsilon}{T_0}\right)^{2/\alpha}A_{\alpha,2}\Gamma\left(3-\frac{2}{\alpha}\right)\nonumber\\
    &\hspace{1em}\times
    \left(1+\frac{3\alpha^2-16\alpha+16}{16\alpha^2}a_2\right)(1-e^2).
\end{align}
For $\alpha=4$, the time evolution is given by
\begin{equation}
    T(t)=T_0 \exp\left(-\frac{t}{\tau_4}\right),
\end{equation}
with
\begin{equation}
    \tau_4^{-1}=2\pi nd^2 \sqrt{\frac{\pi}{m}}A_{4,2}(1-e^2),
\end{equation}
whose dependence on the time is consistent with the previous study \cite{Ernst2002}.
It should be noted that $\tau_4$ is independent of $a_2$ because $S_2$ becomes zero for $\alpha=4$.
For $\alpha>4$, the exponent of the denominator in Eq.\ (\ref{eq:T_t}), $2\alpha/(\alpha-4)$, is larger than $2$ for hard-core gases, which means that the decrease of the temperature for this system is faster than that for inelastic hard-core gases in the later stage of the evolution.

\subsection{Transport coefficients}
New, let us derive the transport coefficients from the Boltzmann equation.
Taking the zeroth, first, and second velocity moments of the Boltzmann equation (\ref{eq:Boltzmann_eq}) yields the balance equations for mass, momentum, and energy, respectively:
\begin{subequations}\label{eq:hydro}
\begin{align}
    &\frac{\partial n}{\partial t}+\bm\nabla \cdot (n \bm{u})=0,\label{eq:hydro1}\\
    &\frac{\partial \bm{u}}{\partial t}+\bm{u}\cdot \bm\nabla \bm{u}+\frac{1}{nm}\bm\nabla \cdot P=0,\label{eq:hydro2}\\
    &\frac{\partial T}{\partial t}+\bm{u}\cdot \bm\nabla T+\frac{2}{3n}(P: \bm\nabla \bm{u}+\bm\nabla \cdot \bm{q})+\zeta T=0.\label{eq:hydro3}
\end{align}
\end{subequations}
Here, $\bm{\nabla}=(\partial/\partial x, \partial/\partial y, \partial/\partial z)$, and the hydrodynamic fields are defined as
\begin{subequations}
\begin{align}
    n&=\int \mathrm{d}\bm{v} f(\bm{v},t),\quad
    \bm{u}=\frac{1}{n}\int \mathrm{d}\bm{v} \bm{v}f(\bm{v},t),\\
    T&=\frac{m}{3n}\int \mathrm{d}\bm{v} V^2 f(\bm{v},t),
\end{align}
\end{subequations}
where $\bm{V}=\bm{v}-\bm{u}$ is the peculiar velocity. 
The pressure tensor and heat flux are microscopically defined by
\begin{equation}
    P_{ij}=m\int \mathrm{d}\bm{v}V_iV_j f(\bm{v},t),\quad
    \bm{q}=\frac{m}{2}\mathrm{d}\bm{v} V^2\bm{V}f(\bm{v},t).
\end{equation}

To derive the constitutive relations and the associated transport coefficients from the Boltzmann equation, we employ the Chapman-Enskog expansion. 
The distribution function is expanded in powers of the hydrodynamic gradients as
\begin{equation}
    f= f^{(0)}+\epsilon f^{(1)}+\cdots,
\end{equation}
where $f^{(0)}$ is the homogeneous solution given by Eq.\ (\ref{eq:f0}), and $\epsilon$ is a formal parameter characterizing the order of the spatial gradients. 
We also expand the time derivative as
\begin{equation}
    \frac{\partial}{\partial t}
    =\frac{\partial^{(0)}}{\partial t}
    +\epsilon \frac{\partial^{(1)}}{\partial t}
    +\cdots.
\end{equation}
At the first order in the hydrodynamic gradients, the pressure tensor and heat flux take the Navier-Stokes forms
\begin{subequations}\label{eq:constitutive1_2}
\begin{align}
    P_{ij}&=p\delta_{ij}-\eta \left(\nabla_i u_j + \nabla_j u_i - \frac{2}{3}\delta_{ij} \bm\nabla\cdot \bm{u}\right),\label{eq:constitutive1}\\
    \bm{q}&= -\kappa \bm\nabla T - \mu \bm\nabla n,\label{eq:constitutive2}
\end{align}
\end{subequations}
where $p=nT$ is the hydrostatic pressure, $\eta$ is the shear viscosity, $\kappa$ is the thermal conductivity, and $\mu$ is the coefficient associated with the contribution of the density gradient to the heat flux.
Equation (\ref{eq:constitutive2}) represents the generalized Fourier law for granular gases.
Unlike ordinary molecular gases, the heat flux contains an additional contribution proportional to $\bm{\nabla} n$.
This term arises from collisional energy dissipation and vanishes in the elastic limit.

Projecting the first-order Chapman-Enskog equation
onto the tensor basis $\tilde{D}_{ij}$~\cite{Brey1998,Garzo1999,Takada2015}, we can obtain the differential equation of the shear viscosity $\eta$ \cite{Brilliantov, Takada2015} as
\begin{equation}
    -\frac{2}{3}T \mu_2 \frac{\partial \eta}{\partial T} - \frac{2}{5}\Omega_\eta^e \eta = \frac{1}{d^2}\sqrt{\frac{mT}{2}},\label{eq:eta_eq}
\end{equation}
where the quantity $\Omega_\eta^e$ is defined as
\begin{align}
    \Omega_\eta^e &\equiv \int \mathrm{d}\bm{c}_1 \int \mathrm{d}\bm{c}_2 \int \mathrm{d}\hat{\bm{k}} \tilde{\sigma}(\chi,c_{12}) c_{12}\phi(c_1)\phi(c_2)\nonumber\\
    &\hspace{1em}\times\left[1+a_2 S_2(c_1^2)\right] \tilde{D}_{ij}(\bm{c}_2) \Delta \left[\tilde{D}_{ij}(\bm{c}_1)+\tilde{D}_{ij}(\bm{c}_2)\right]\nonumber\\
    &=-\sqrt{2\pi}\left(\frac{\varepsilon}{T}\right)^{2/\alpha}\Gamma\left(4-\frac{2}{\alpha}\right)
    \left(\omega_{\eta,1}+a_2\omega_{\eta,2}\right),\label{eq:Omega_eta}
\end{align}
where $\omega_{\eta,1}$ and $\omega_{\eta,2}$ are given by Eqs.\ (\ref{eq:omega_eta1_eta2}), respectively (the detailed derivation is shown in Appendix \ref{sec:BasicIntegral}).
The effective cross section scales as
\begin{equation}
    \sigma_{\rm eff}
    \propto
    \left(\frac{\varepsilon}{T}\right)^{2/\alpha}.
\end{equation}
Since the shear viscosity is proportional to
$\sqrt{T}/\sigma_{\rm eff}$, 
\begin{equation}
    \eta
    \propto T^{1/2+2/\alpha}
    = T^{(\alpha+4)/(2\alpha)}.
\end{equation}
Therefore,
\begin{equation}
    T \frac{\partial \eta}{\partial T}
    = \frac{\alpha+4}{2\alpha}\eta
\end{equation}
is satisfied.
\begin{figure}[htbp]
    \begin{center}
	\includegraphics[width=85mm]{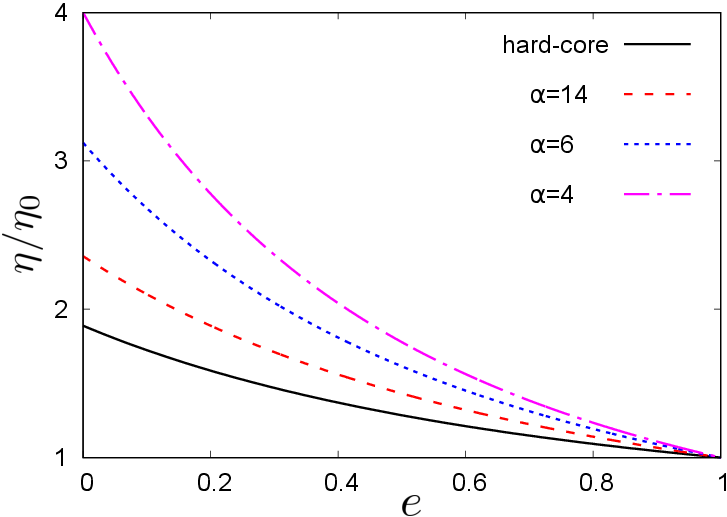}
    \end{center}
    \caption{Plot of the shear viscosity against the restitution coefficient $e$ given by Eq.\ (\ref{eq:a2}) for $\alpha=14$ (red dashed line), $6$ (blue dotted line), and $4$ (pink dot-dashed line). 
    Here, $\eta_0=\eta(1)$ is the viscosity for elastic gases. 
    The solid line shows that for inelastic hard-core gases.}
    \label{fig:eta}
\end{figure}
Substituting this relationship into Eq.\ (\ref{eq:eta_eq}), the explicit form of the shear viscosity is given by
\begin{align}
    \eta(e) &= -\frac{1}{d^2}\sqrt{\frac{mT}{2}}\left(\frac{\alpha+4}{3\alpha}\mu_2 + \frac{2}{5}\Omega_\eta^e\right)^{-1}\nonumber\\
    &=\eta_0\frac{48(3\alpha-2)(A_{\alpha,2}-A_{\alpha,4})}{6(3\alpha-2)\omega_{\eta,1}-5(\alpha+4)S_1}\nonumber\\
    &\hspace{1em}\times\left[1-\frac{6(3\alpha-2)\omega_{\eta,2}-5(\alpha+4)S_2}{6(3\alpha-2)\omega_{\eta,1}-5(\alpha+4)S_1}a_2\right],
    \label{eq:eta}
\end{align}
with the elastic shear viscosity
\begin{align}
    \eta_0=\frac{5}{\displaystyle 32(A_{\alpha,2}-A_{\alpha,4})\Gamma\left(4-\frac{2}{\alpha}\right)d^2}
    \left(\frac{m}{\pi}\right)^{1/2}\frac{T^{(\alpha+4)/(2\alpha)}}{\varepsilon^{2/\alpha}},
\end{align}
under the linear approximation with respect to $a_2$.
It is noted that $\eta_0$ is consistent with the previous studies for elastic gases \cite{Chapman,Hirschfelder}.
We also note that Eq.\ (\ref{eq:eta}) reduces to that for hard-core gases for $\alpha\to \infty$.
Figure \ref{fig:eta} shows the restitution coefficient dependence of the shear viscosity.
Similar to the behavior of $a_2$, the deviation from the elastic viscosity becomes smaller for larger $\alpha$ and the shear viscosity is monotonically decreasing function with respect to the restitution coefficient $e$.
For $\alpha=4$, the reduced shear viscosity $\eta/\eta_0$ is about twice larger than that for hard-core gases.

\begin{figure}[htbp]
    \begin{center}
        \includegraphics[width=85mm]{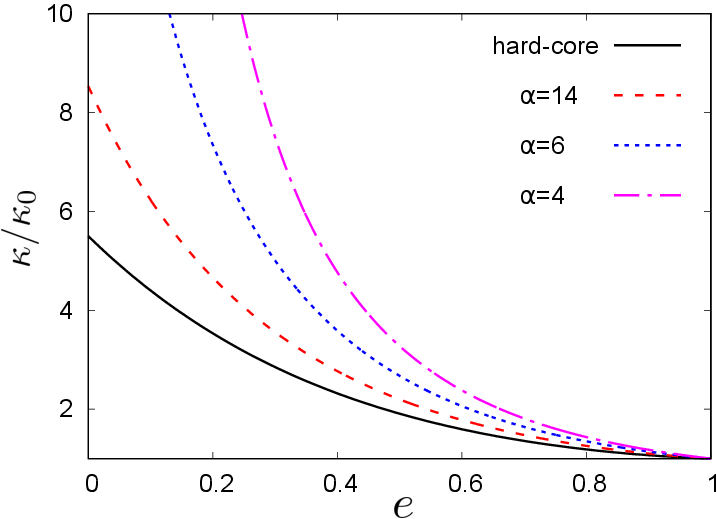}
    \end{center}
    \caption{Plot of the thermal conductivity against the restitution coefficient $e$ given by Eq.\ (\ref{eq:a2}) for $\alpha=14$ (red dashed line), $6$ (blue dotted line), and $4$ (pink dot-dashed line).
    Here, $\kappa_0=\kappa(1)$ is the thermal conductivity for elastic gases. 
    The solid line shows that for inelastic hard-core gases.}
    \label{fig:kappa}
\end{figure}

\begin{figure}[htbp]
    \begin{center}
        \includegraphics[width=85mm]{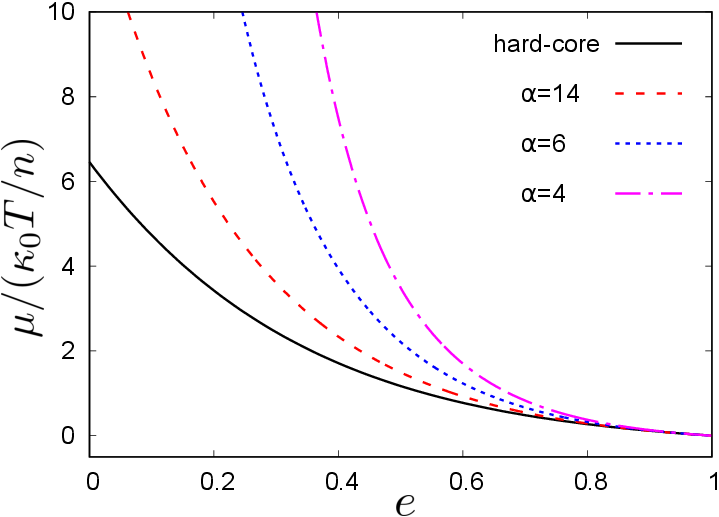}
    \end{center}
    \caption{Plot of the coefficient $\mu$ against the restitution coefficient $e$ given by Eq.\ (\ref{eq:a2}) for $\alpha=14$ (red dashed line), $6$ (blue dotted line), and $4$ (pink dot-dashed line).
    The solid line shows that for inelastic hard-core gases.}
    \label{fig:mu}
\end{figure}

Next, let us derive the expressions of the thermal conductivity and the coefficient $\mu$.
We define the quantity $\Omega_\kappa^e$ as
\begin{align}
    \Omega_\kappa^e &= \int \mathrm{d}\bm{c}_1 \int \mathrm{d}\bm{c}_2 \int \mathrm{d}\hat{\bm{k}} \tilde{\sigma}(\chi,c_{12})c_{12}\phi(c_1)\phi(c_2)\nonumber\\
	&\hspace{1em}\times\left[1+a_2S_2(c_1^2)\right]\tilde{\bm{S}}(\bm{c}_2)\cdot \Delta\left[\tilde{\bm{S}}(\bm{c}_1)+\tilde{\bm{S}}(\bm{c}_2)\right]\nonumber\\
	&=-\sqrt{2\pi}\left(\frac{\varepsilon}{T}\right)^{2/\alpha}\Gamma\left(4-\frac{2}{\alpha}\right)\left(\omega_{\kappa,1}+a_2\omega_{\kappa,2}\right),
\end{align}
where $\omega_{\kappa,1}$ and $\omega_{\kappa,2}$ are given by Eqs.\ (\ref{eq:omega_kappa1_kappa2}), respectively (the detailed derivation is shown in Appendix \ref{sec:BasicIntegral}).
Similarly, projecting onto the vector basis
$\bm{S}$, we can obtain the differential equations of the thermal conductivity $\kappa$ and the coefficient $\mu$ \cite{Brilliantov, Takada2015, Takada2022} as
\begin{subequations}\label{eq:kappa_eq_mu_eq}
\begin{align}
    &\frac{\partial}{\partial T}\left(4\mu_2 \kappa T^{3/2}\right) +\frac{8}{5}\kappa T^{1/2}\Omega_\kappa^e = -\frac{15}{2}\frac{T}{d^2}\sqrt{\frac{2}{m}}\left(1+2a_2\right),\label{eq:kappa_eq}\\
    &-4T\mu_2\frac{\partial\mu}{\partial T}-\frac{8}{5}\Omega_\kappa^e \mu = \frac{4}{n}T\mu_2 \kappa + a_2\frac{15}{2nd^2}\sqrt{\frac{2T^3}{m}},\label{eq:mu_eq}
\end{align}
\end{subequations}
respectively. 
From the same discussion about the shear viscosity, the thermal conductivity and the coefficient $\mu$ should scale as $\kappa\sim T^{(\alpha+4)/(2\alpha)}$ and $\mu\sim T^{(3\alpha+4)/(2\alpha)}$, respectively.
From Eqs.\ (\ref{eq:kappa_eq_mu_eq}), we can obtain the explicit forms of them as
\begin{align}
    \kappa(e)&=-\frac{15}{16d^2}\sqrt{\frac{2T}{m}}(1+2a_2)\left(\mu_2+\frac{1}{5}\Omega_\kappa^e\right)^{-1}\nonumber\\
    &=\kappa_0 \frac{8(3\alpha-2)(A_{\alpha,2}-A_{\alpha,4})}{(3\alpha-2)\omega_{\kappa,1}-5\alpha S_1}\nonumber\\
    &\hspace{1em}\times\left[1+\frac{(3\alpha-2)(2\omega_{\kappa,1}-\omega_{\kappa,2})-5\alpha (2S_1-S_2)}{(3\alpha-2)\omega_{\kappa,1}-5\alpha S_1}a_2\right],\label{eq:kappa}
\end{align}
and
\begin{align}
    &\mu(e)\nonumber\\
    &=-\left(\frac{2}{n}T\mu_2\kappa+a_2\frac{15}{4nd^2}\sqrt{\frac{2T^3}{m}}\right)\left(\frac{3\alpha+4}{\alpha}\mu_2+\frac{4}{5}\Omega_\kappa^e\right)^{-1}\nonumber\\
    &=\frac{T}{n}\kappa_0 \frac{80\alpha(3\alpha-2)(A_{\alpha,2}-A_{\alpha,4})S_1}{[(3\alpha-2)\omega_{\kappa,1}-5\alpha S_1][4(3\alpha-2)\omega_{\kappa,1}-5(3\alpha+4)S_1]}\nonumber\\
    &\hspace{1em}\times\left\{1+\left[\frac{S_2}{S_1}+\frac{(3\alpha-2)(2\omega_{\kappa,1}-\omega_{\kappa,2})-5\alpha (2S_1-S_2)}{(3\alpha-2)\omega_{\kappa,1}-5\alpha S_1}\right.\right.\nonumber\\
    &\hspace{2.5em}\left.\left.+\frac{9}{20}\frac{(3\alpha-2)\omega_{\kappa,1}-5\alpha S_1}{\alpha S_1}\right.\right.\nonumber\\
    &\hspace{2.5em}\left.\left.
    -\frac{4(3\alpha-2)\omega_{\kappa,2}-5(3\alpha+4)S_2}{4(3\alpha-2)\omega_{\kappa,1}-5(3\alpha+4)S_1}\right]a_2\right\},\label{eq:mu}
\end{align}
respectively, under the linear approximation with respect to $a_2$, where the elastic thermal conductivity is defined as
\begin{align}
    \kappa_0&=
    \frac{75}{\displaystyle 144(A_{\alpha,2}-A_{\alpha,4})\Gamma\left(4-\frac{2}{\alpha}\right)d^2}\nonumber\\
    &\hspace{1em}\times\left(\frac{1}{\pi m}\right)^{1/2}\frac{T^{(\alpha+4)/(2\alpha)}}{\varepsilon^{2/\alpha}}.\label{eq:kappa_elastic}
\end{align}
It is noted that Eqs.\ (\ref{eq:kappa}) and (\ref{eq:mu}) reduce to those for hard-core gases for $\alpha\to \infty$, respectively, and Eq.\ (\ref{eq:kappa_elastic}) is consistent with the previous studies \cite{Chapman,Hirschfelder}.
As expected, $\mu$ vanishes in the elastic limit
$e \to 1$ for arbitrary $\alpha$.
Figures \ref{fig:kappa} and \ref{fig:mu} show the restitution coefficient dependence of the thermal conductivity $\kappa$ and the coefficient $\mu$, respectively.
Similar to the behavior of $\eta$, the deviations becomes smaller for larger $\alpha$, while these deviations are larger than that of the reduced shear viscosity.
The divergence of $\kappa$ and $\mu$ originates from the
vanishing denominator of Eqs.~(\ref{eq:kappa}) and (\ref{eq:mu}), which corresponds to a balance between collisional cooling and collisional transfer contributions.

\subsection{Linear stability analysis}
In this subsection, let us perform linear stability analysis for homogeneous cooling state of this system.
Using the constitutive relations (\ref{eq:constitutive1_2}), we can rewrite a set of hydrodynamic equations (\ref{eq:hydro}) as
\begin{subequations}
\begin{align}
    &\frac{\partial n}{\partial t}+\bm\nabla \cdot (n\bm{u})=0,\\
    &\frac{\partial \bm{u}}{\partial t}+\bm{u}\cdot \bm\nabla \bm{u}=-\frac{1}{nm}\bm\nabla p + \frac{\eta}{nm}\left[\nabla^2 \bm{u}+\frac{1}{3}\bm\nabla (\bm\nabla \cdot \bm{u})\right],\\
    &\frac{\partial T}{\partial t}+\bm{u}\cdot \bm\nabla T\nonumber\\
    &=-\zeta T +\frac{2}{3n}\left(\kappa \nabla^2T + \mu \nabla^2 n\right)
    -\frac{2}{3n}p (\bm\nabla \cdot \bm{u})\nonumber\\
    &\hspace{1em}+\frac{2}{3n}\eta \left[(\nabla_i u_j)(\nabla_j u_i)+(\nabla_j u_i)(\nabla_i u_j)-\frac{2}{3}(\bm\nabla \cdot \bm{u})^2\right].
\end{align}
\end{subequations}
We linearize these equations around the homogeneous cooling state, and only consider the linear term with respect to the density $\rho\equiv \delta n/n_{\rm H}$, $\bm{w}\equiv \bm{u}/v_{\rm T}$, and $\theta\equiv \delta T/T_{\rm H}$, where $\delta n$ and $\delta T$ are the fluctuations around the homogeneous state of the density and the temperature, respectively, and $n_{\rm H}$ and $T_{\rm H}$ are the density and the temperature of homogeneous system, respectively.
We also define dimensionless time and space variables $\tau$ and $\hat{\bm{r}}$ as $\tau =\int_0^t \mathrm{d}t^\prime \nu_{\rm H}(t^\prime)$ and $\hat{\bm{r}}=(2\nu_{\rm H}(t)/v_{\rm T}(t))\bm{r}$, respectively, with the collision frequency $\nu_{\rm H}=p/\eta(1)$.
Let us define the reduced quantities $\eta^*\equiv \eta/\eta_0$, $\kappa^*\equiv \kappa/\kappa_0$, $\mu^*\equiv\mu/(\kappa_0T/n)$, and $\zeta^*\equiv \zeta(e)\eta_0/(nT)$.
We also define the Fourier component of the quantity $A$ as
\begin{equation}
    A_{\bm k}(\tau)= \int \mathrm{d}\hat{\bm{r}} A(\hat{\bm{r}},\tau) 
    \mathrm{e}^{-\mathrm{i} \bm{k}\cdot \hat{\bm{r}}},
\end{equation}
with $\mathrm{i}=\sqrt{-1}$.
After the linearization around the homogeneous cooling state and the Fourier transformation, we can rewrite the set of hydrodynamic equations as
\begin{subequations}
\begin{align}
    \frac{\partial}{\partial \tau}\rho_{\bm k} 
    &= -\mathrm{i}k w_{\bm{k}\parallel},\label{eq:LS_eq1}\\
    \frac{\partial}{\partial \tau}w_{\bm{k}\parallel} 
    &= -\frac{1}{2}\mathrm{i}k \rho_{\bm k} 
    +\left(\frac{1}{4}\zeta^*-\frac{4}{3}\eta^* k^2\right) w_{\bm{k}\parallel} -\frac{1}{2}\mathrm{i}k \theta_{\bm k},\label{eq:LS_eq2}\\
    \frac{\partial}{\partial \tau}w_{\bm{k}\perp}
    &=\left(\frac{1}{4}\zeta^* - \eta^* k^2\right)w_{\bm{k}\perp},\label{eq:LS_eq3}\\
    \frac{\partial}{\partial \tau}\theta_{\bm k}&=\left(-\frac{1}{2}\zeta^* -\frac{5}{2}\mu^* k^2\right)\rho_{\bm k}
    -\frac{2}{3}\mathrm{i}k w_{\bm{k}\parallel} \nonumber\\
    &\hspace{1em}+ \left(-\frac{\alpha-4}{4\alpha}\zeta^*-\frac{5}{2}\kappa^* k^2\right)\theta_{\bm{k}}.\label{eq:LS_eq4}
\end{align}
\end{subequations}
Equation (\ref{eq:LS_eq3}) can be solved as
\begin{equation}
    w_{\bm{k}\perp}(\tau)=w_{\bm{k}\perp}(0)\exp\left[\left(\frac{1}{4}\zeta^*-\eta^*k^2\right)\tau\right].
\end{equation}
This is a shear mode and the threshold is given by
\begin{equation}
    k_\perp = \frac{1}{2}\sqrt{\frac{\zeta^*}{\eta^*}}.\label{eq:shear_mode}
\end{equation}
Other three eigenmodes are given by the solution of the following eigenequation from Eqs.\ (\ref{eq:LS_eq1}), (\ref{eq:LS_eq2}), and (\ref{eq:LS_eq4}):
\begin{align}
    &\lambda^3+\left[-\frac{1}{\alpha}\zeta^* + \left(\frac{4}{3}\eta^*+\frac{5}{2}\kappa^*\right)k^2\right]\lambda^2\nonumber\\
    &+\left[-\frac{\alpha-4}{16\alpha}\zeta^{*2}+\left(\frac{5}{6}+\frac{\alpha-4}{3\alpha}\zeta^*\eta^*-\frac{5}{8}\zeta^*\kappa^*\right)k^2\right.\nonumber\\
    &\hspace{1em}\left.+\frac{10}{3}\eta^*\kappa^* k^4\right]\lambda
    +\left[-\frac{\alpha+4}{8\alpha}\zeta^*k^2+\frac{5}{4}(\kappa^*-\mu^*)k^4\right]=0.\label{eq:eigen_eq}
\end{align}
This determines a heat mode and two sound modes.
The threshold for the heat mode, where $\lambda_{\rm H}=0$ is satisfied, can be determined from Eq.\ (\ref{eq:eigen_eq}) as
\begin{equation}
    k_{\rm H}=\sqrt{\frac{(\alpha+4)\zeta^*}{10\alpha(\kappa^*-\mu^*)}}.\label{eq:heat_mode}
\end{equation}
\begin{figure}[htbp]
    \begin{center}
	\includegraphics[width=85mm]{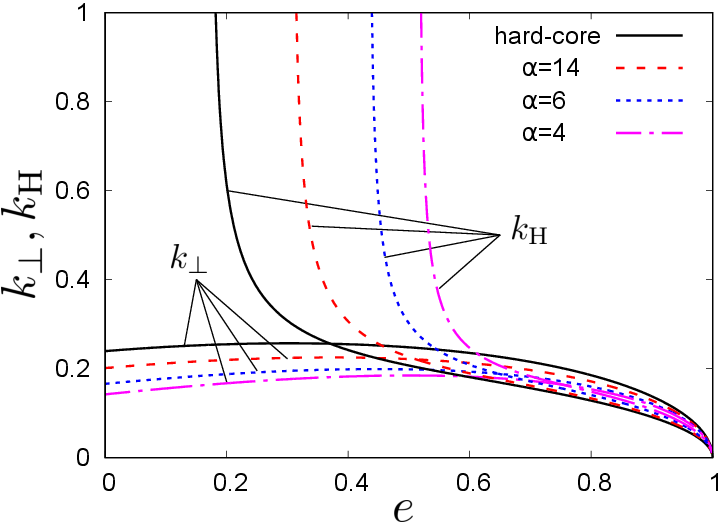}
    \end{center}
    \caption{Plot of the thresholds for the shear mode and the heat mode against the restitution coefficient $e$ for $\alpha=14$ (red dashed line), $6$ (blue dotted line), and $4$ (pink dot-dashed line).
    The solid lines show those for inelastic hard-core gases.}
    \label{fig:threshold}
\end{figure}
Figure \ref{fig:threshold} shows the restitution coefficient dependence of the thresholds for the shear mode (\ref{eq:shear_mode}) and the heat mode (\ref{eq:heat_mode}).
The threshold for the shear mode becomes larger for harder system.
On the other hand, the threshold for the heat mode has opposite dependence on the softness of the potential, where the threshold becomes larger for softer potential.
This is because the larger $\mu^*$ destabilize the system as shown in Eq.\ (\ref{eq:heat_mode}).
In addition, it diverges for small $e$, where $\kappa^*=\mu^*$ is satisfied in Eq.\ (\ref{eq:heat_mode}), that is $e_{\rm c}=0.434$ for $\alpha=6$, which was also reported for hard-core system \cite{Brey1996}.
This divergent point $e_{\rm c}$ is a decreasing function of $\alpha$, and $e_{\rm c}=0.516$, $0.307$, and $0.171$ for $\alpha=4$, $14$, and $\infty$, respectively.
\begin{figure}[htbp]
    \begin{center}
	\includegraphics[width=85mm]{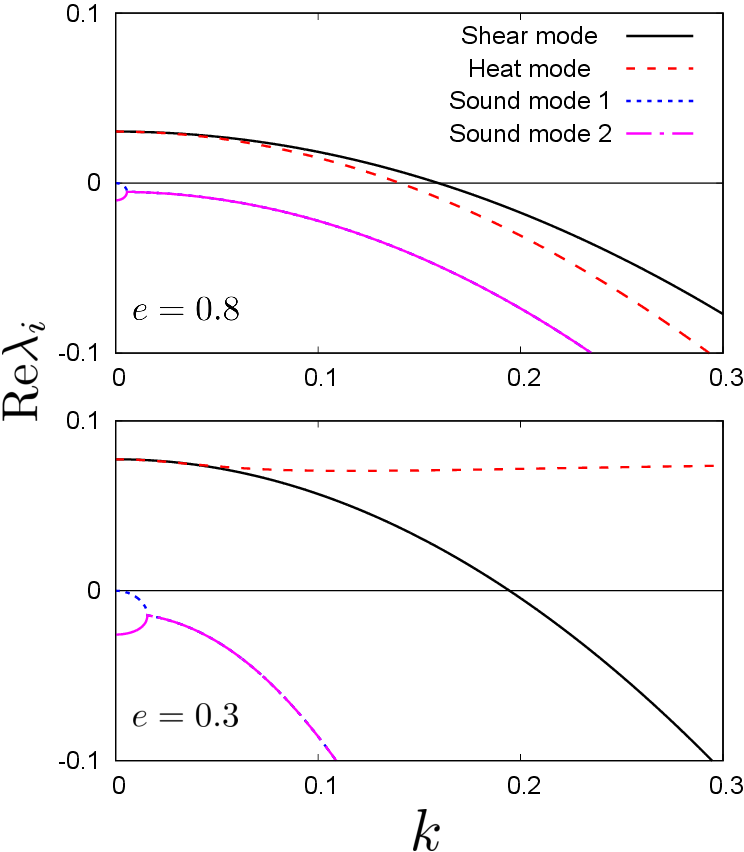}
    \end{center}
    \caption{Plot of the real part of each mode against the wave number for $e=0.8$ (top) and $e=0.3$ (bottom) for $\alpha=6$.}
    \label{fig:eigen}
\end{figure}
Figure \ref{fig:eigen} shows the wave number $k$ dependence of each mode for $e=0.8$ and $0.3$ and $\alpha=6$.
For $e=0.8$, the shear and heat modes are monotonically decreasing function of $k$, and become negative above the thresholds given by Eqs.\ (\ref{eq:shear_mode}) and (\ref{eq:heat_mode}), respectively as shown in Fig.\ \ref{fig:eigen} (top).
For $e=0.3$, on the other hand, the heat mode is always positive for any wave number as shown in Fig.\ \ref{fig:eigen} (bottom), which means the divergence of the heat mode for small $e$, while the shear mode is still decreasing function.

\section{Discussion}\label{sec:discussion}
Let us compare our results for $\alpha=4$ with those reported in the previous studies.
In Ref.\ \cite{Khalil2014}, the generalized inelastic Maxwell model was studied, where the temperature dependence of the collision frequency is $T^\gamma$.
They obtained the explicit forms of the transport coefficients for this system, and for example, the temperature dependence of the shear viscosity is given by $T^{1-\gamma}$.
For inelastic Maxwell model, the collision frequency is independent of the temperature, which means $\gamma=0$.
Putting $\gamma=0$ into the shear viscosity derived in Ref.\ \cite{Khalil2014}, its temperature dependence becomes linear, which is consistent with our result (\ref{eq:eta}) for $\alpha=4$.
It is also noted that if we choose $\gamma=(\alpha-4)/2\alpha$ in Ref.\ \cite{Khalil2014}, the temperature dependence of the transport coefficients is qualitatively consistent with that of our results for arbitrary $\alpha$, while the quantitative agreement is poor because of the ignorance of the scattering processes.

\begin{figure}[htbp]
    \begin{center}
	\includegraphics[width=85mm]{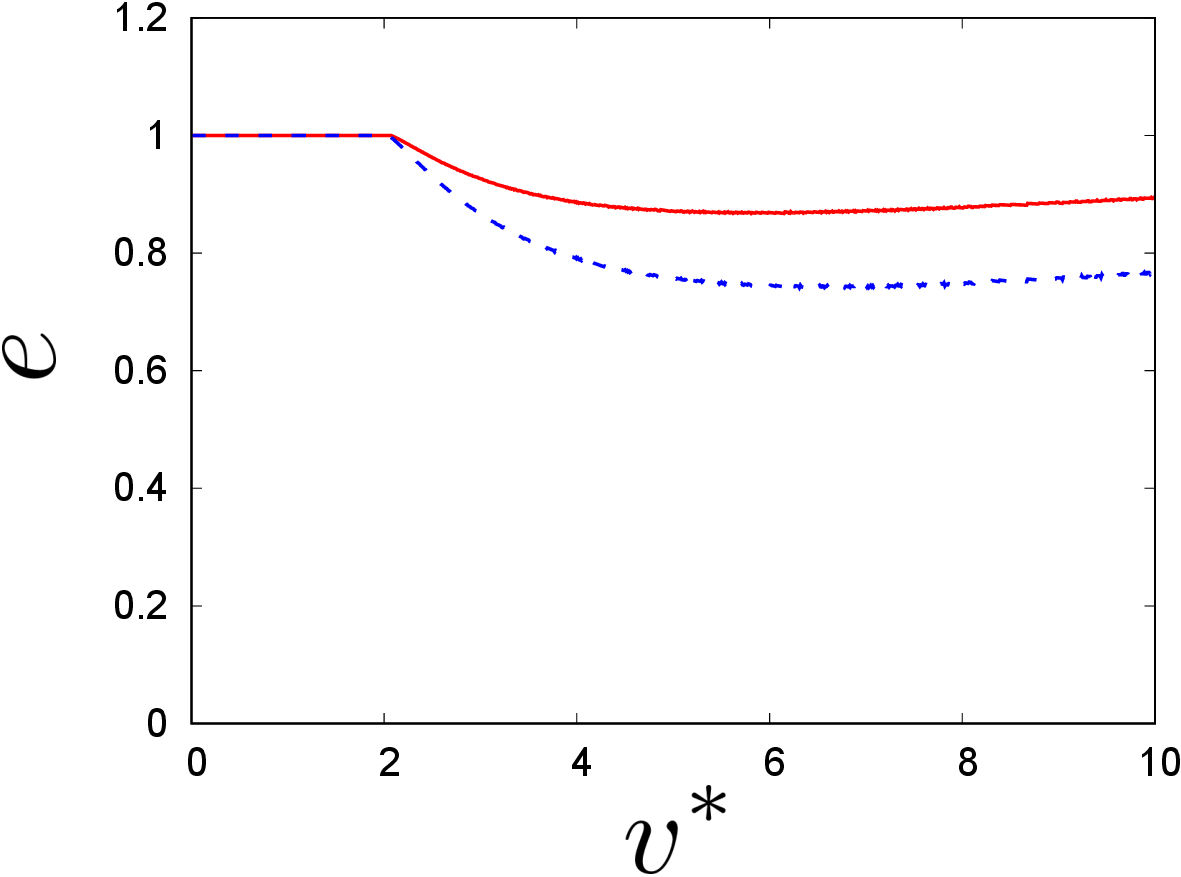}
    \end{center}
    \caption{Velocity dependence of the restitution coefficient for the inverse power law potential with the damping term (\ref{eq:F_vis}) for $\zeta_{\rm vis}=\sqrt{d^2/m\varepsilon}$ (red solid line) and $2\sqrt{d^2/m\varepsilon}$ (blue dashed line), where $v^*\equiv v\sqrt{m/\varepsilon}$.}
    \label{fig:e_v}
\end{figure}
For $\alpha<4$, the time evolution of the temperature changes drastically.
In this case, the characteristic time $\tau_\alpha$ and the exponent of the denominator in Eq.\ (\ref{eq:T_t}) becomes negative, 
which means that our theory is only valid for $t\le -\tau_\alpha(>0)$ because the temperature becomes zero at $t=-\tau_{\alpha}$ and thereafter all particles stop.
This singular behavior comes from the choice of the model of inelasticity.
As mentioned in Sec.\ \ref{sec:Boltzmann}, our model is oversimplified model, because we ignore the velocity and the angle dependence of the restitution coefficient.
In actual systems, the energy dissipation is characterized by core collisions or deformations of the surface \cite{Dufty2008}.
For example, this can be modeled by the following force $\bm{F}_i$ acting on the $i$-th particle:
\begin{equation}
    \bm{F}_i= -\sum_{j\neq i} \bm\nabla_i U(|\bm{r}_{ij}|)-\zeta_{\rm vis} \sum_{j\neq i}\Theta(d-|\bm{r}_{ij}|)(\bm{v}_{ij}\cdot \hat{\bm{r}}_{ij})\hat{\bm{r}}_{ij},\label{eq:F_vis}
\end{equation}
instead of the constant restitution coefficient (\ref{eq:pre_post}), where $i$ and $j$ label the particles, $\Theta(r)$ is a step function, $\bm{r}_{ij}=\bm{r}_i-\bm{r}_j$ and $\bm{v}_{ij}=\bm{v}_i-\bm{v}_j$ are the relative position and velocity, respectively, $\hat{\bm{r}}_{ij}=\bm{r}_{ij}/|\bm{r}_{ij}|$, and $\zeta_{\rm vis}$ is the dissipation rate.
Figure \ref{fig:e_v} shows the initial velocity dependence of the restitution coefficient when two particles collide head-on.
In this case, the energy dissipation happens only when the particle overcomes the repulsive force $\sim r^{-(\alpha+1)}$ of the other particle and there is a threshold for the speed of the particles.
This means that when the temperature is sufficiently small, almost all particles cannot approach because of the strong repulsive forces and does not lose energy.
Thus, the oversimplification of the inelasticity results in the singular behavior of the temperature evolution.
As shown in Fig.\ \ref{fig:e_v}, however, our simple model (\ref{eq:pre_post}) can be regarded as a good approximation when the average velocity of the particles are sufficiently high.
Quantitatively, we need the modification of the treatment, but this is a future work.

\section{Summary}\label{sec:summary}
In this paper, we have derived the transport coefficients of granular gases having an inverse power law potential. 
We have investigated the homogeneous cooling state and found that the decrease of the temperature is proportional to $t^{-2\alpha/(\alpha-4)}$ for $\alpha\neq4$ in the long time limit and exponential to the time for $\alpha=4$. 
We have also derived the transport coefficients using the Chapman-Enskog theory and obtained the explicit temperature dependence of them. 
The linear stability analysis has shown the divergence of the threshold for the heat mode for small restitution coefficient.
We have also found that the behaviors of the thresholds for the shear and heat modes are opposite with respect to the softness of the potential.

\section*{Acknowledgments}
The author thanks A.\ Santos and T.\ Araki for useful comments.
Numerical computation in this work was partially carried out at the Yukawa Institute Computer Facility, Kyoto University.

\appendix
\section{Scattering processes}\label{sec:scattering}
In this Appendix, we briefly summarize the scattering processes and show its explicit form for some exponents.
Let us consider a scattering process in which two particles approach from far away with relative velocity $\bm{v}$ and leave with the relative velocity $\bm{v}^\prime$ after the scattering in the frame that the target is stationary.
The incident angle $\theta$ between $\bm{v}$ and the normal unit vector $\hat{\bm{k}}$ at the closest distance $r=r_{\rm min}$ between colliding particles is given by
\begin{align}
    \theta_\alpha 
    &= b \int_0^{u_0} \frac{\mathrm{d}u}{\displaystyle\sqrt{1-b^2 u^2 -\frac{4}{m v^2} U(1/u)}}\nonumber\\
    &=\int_0^{x_0} \frac{\mathrm{d}x}{\displaystyle \sqrt{1-x^2-\left(\frac{x}{\tilde{b}}\right)^\alpha}},\label{eq:theta_def}
\end{align}
where $\tilde{b}=(mv^2/4\varepsilon)^{1/\alpha}b/d$ and $x_0=b u_0$ with $u_0=1/r_{\rm min}$ which is the smaller one of the positive solutions of the inner of the denominator in Eq.\ (\ref{eq:theta_def}).
It should be noted that $\theta_\alpha$ is only a function of $\tilde{b}$ and we can evaluate Eq.\ (\ref{eq:theta_def}) for arbitrary exponent $\alpha$.
We also note that Eq.\ (\ref{eq:theta_def}) is analytically solvable for some $\alpha$.
For $\alpha=4$, Eq.\ (\ref{eq:theta_def}) reduces to
\begin{equation}
    \theta_4 = \frac{\tilde{b}}{(\tilde{b}^4+4)^{1/4}} K\left(\frac{1}{2}-\frac{\tilde{b}^2}{2\sqrt{\tilde{b}^4+4}}\right),
\end{equation} 
with the complete elliptic integral of the first kind $K(k)$ \cite{Abramowitz}.
For $\alpha=6$, Eq.\ (\ref{eq:theta_def}) can be rewritten in terms of $y=1/x^2-1/3$ as
\begin{align}
    \theta_6&=\int_{1/x_0^2-1/3}^\infty \frac{\mathrm{d}y}{\displaystyle \sqrt{4y^3-\frac{4}{3}y-\frac{8}{27}-\frac{4}{\tilde{b}^6}}}\nonumber\\
    &\equiv \wp^{-1}(e_2;g_2,g_3)=\omega_2,
\end{align}
where $\wp^{-1}(z;g_2, g_3)$ is the inverse Weierstrass' $\wp$ function \cite{Abramowitz} with the two invariants $g_2=4/3$ and $g_3=8/27+4/\tilde{b}^6$, $e_2\equiv 1/x_0^2-1/3$, and $\omega_2$ is the half period relating to $e_2$.
For $n=14$, on the other hand, we should numerically solve Eq.\ (\ref{eq:theta_def}) because this integral cannot be solved analytically.

\section{Basic integral}\label{sec:BasicIntegral}
In this Appendix, let us show the detailed derivation of the transport coefficients.
First, we introduce the general integral to make our calculations in Sec.\ \ref{sec:Boltzmann} simpler.
We define the general integral $J_{k,l,m,n,p}$ as
\begin{align}
    J_{k,l,m,n,p} &\equiv 
    \int \mathrm{d}\bm{C} \int \mathrm{d}\bm{c}_{12} \int \mathrm{d}\hat{\bm{k}} \tilde{\sigma}
    (\chi,c_{12}) c_{12} \phi(C) \phi(c_{12})\nonumber\\
    &\hspace{1em}\times C^k c_{12}^l (\bm{C}\cdot \bm{c}_{12})^m 
    (\bm{C}\cdot \hat{\bm{k}})^n (\bm{c}_{12}\cdot \hat{\bm{k}})^p,\label{eq:int_J}
\end{align}
with $\bm{C}=(\bm{c}_1+\bm{c}_2)/2$.
Here, we rewrite the integration with respect to $\hat{\bm{k}}$ as \cite{Cercignani}
\begin{align}
    \int \mathrm{d}\hat{\bm{k}}\tilde{\sigma}(\chi,c_{12})c_{12}
    &=\int_0^\pi \mathrm{d}\chi \int_0^{2\pi} \mathrm{d}\phi\sin\chi c_{12}\frac{b^*}{\sin\chi}\left|\frac{\partial b^*}{\partial \chi}\right|\nonumber\\
    &=2\pi \int_0^\infty \mathrm{d}b^* b^* c_{12}\nonumber\\
    &=2\pi \cdot 2^{-1/\alpha} \left(\frac{\varepsilon}{T}\right)^{2/\alpha}
    \int_0^\infty \mathrm{d}\tilde{b} \hspace{0.2em}\tilde{b}c_{12}^{1-4/\alpha},\label{eq:int_dk}
\end{align}
where $b^*\equiv b/d$.
Using Eq.\ (\ref{eq:int_dk}) and after some manipulation, Eq.\ (\ref{eq:int_J}) can be rewritten as
\begin{align}
    J_{k,l,m,n,p} 
    &=\left(\frac{\varepsilon}{T}\right)^{2/\alpha}2^{-(k-l+n-p-1)/2} \pi^{-1/2}\nonumber\\
    &\hspace{1em}\times 
    \frac{\displaystyle \Gamma\left(\frac{k+m+n+3}{2}\right)\Gamma\left(\frac{l+m+p+4}{2}-\frac{2}{\alpha}\right)}
    {\displaystyle \Gamma\left(\frac{m+n+3}{2}\right)}\nonumber\\
    &\hspace{1em}\times \sum_{j=0}^n \dbinom{n}{j} \left[1+(-1)^j\right]
    \left[1+(-1)^{m+n-j}\right]\nonumber\\
    &\hspace{1em}\times 
    \Gamma\left(\frac{1+j}{2}\right)\Gamma\left(\frac{m+n+1-j}{2}\right)\nonumber\\
    &\hspace{1em}\times 
    \int_0^\infty \mathrm{d}\tilde{b} \hspace{0.2em}\tilde{b}\cos^{n+p-j}\theta_\alpha \left(1-\cos^2 \theta_\alpha\right)^{j/2}.\label{eq:def_J}
\end{align}
In this paper, we only focus on the results for $n=0$, $1$ and $2$.
In each case, this integral reduces to
\begin{subequations}
\begin{align}
    J_{k,l,m,0,p}
    &= \left(\frac{\varepsilon}{T}\right)^{2/\alpha}\frac{2^{-(k-l-p-5)/2}}{m+1}\left[1+(-1)^m\right]A_{\alpha,p}\nonumber\\
    &\hspace{1em}\times 
    \Gamma \left(\frac{k+m+3}{2}\right)\Gamma \left(\frac{l+m+p+4}{2}-\frac{2}{\alpha}\right),\\
    J_{k,l,m,1,p}
    &= \left(\frac{\varepsilon}{T}\right)^{2/\alpha}\frac{2^{-(k-l-p-4)/2}}{m+2}\left[1-(-1)^m\right]A_{\alpha,p+1}\nonumber\\
    &\hspace{1em}\times 
    \Gamma \left(\frac{k+m+4}{2}\right)\Gamma \left(\frac{l+m+p+4}{2}-\frac{2}{\alpha}\right),\\
    J_{k,l,m,2,p}
    &= \left(\frac{\varepsilon}{T}\right)^{2/\alpha}\frac{2^{-(k-l-p-3)/2}}{(m+1)(m+3)}\left[1+(-1)^m\right]\nonumber\\
    &\hspace{1em}\times 
    \Gamma \left(\frac{k+m+5}{2}\right)\Gamma \left(\frac{l+m+p+4}{2}-\frac{2}{\alpha}\right)\nonumber\\
    &\hspace{1em}\times \left(A_{\alpha,p}+m A_{\alpha,p+2}\right),
\end{align}
\end{subequations}
respectively.
Using the distribution function (\ref{eq:f0}), we can rewrite Eq.\ (\ref{eq:mu_n}) as
\begin{align}
    \mu_n &=-\frac{1}{2}\int \mathrm{d}\bm{c}_1 \int \mathrm{d}\bm{c}_2 \int \mathrm{d}\hat{\bm{k}}\tilde{\sigma}(\chi,c_{12}) c_{12}\phi(c_1)\phi(c_2)\nonumber\\
    &\hspace{1em}\times 
    \left[1+a_2 \left(S_2(c_1^2)+S_2(c_2^2)\right)\right] \Delta[c_1^n+c_2^n].\label{eq:mu_n_}
\end{align}
For $n=2$ and $4$, $\Delta[c_1^n+c_2^n]$ becomes
\begin{subequations}
\begin{align}
    &\Delta[c_1^2+c_2^2] = -\frac{1}{2}(1-e^2)(\bm{c}_{12}\cdot \hat{\bm{k}})^2,\label{eq:Delta_2}\\
    &\Delta[c_1^4+c_2^4] \nonumber\\
    &= 2(1+e)^2(\bm{C}\cdot \hat{\bm{k}})^2(\bm{c}_{12}\cdot \hat{\bm{k}})^2
    +\frac{1}{8}(1-e^2)^2 (\bm{c}_{12}\cdot \hat{\bm{k}})^4 \nonumber\\
    &\hspace{1em}\times -\frac{1}{4}(1-e^2)^2c_{12}^2 (\bm{c}_{12}\cdot \hat{\bm{k}})^2
    -(1-e^2)C^2(\bm{c}_{12}\cdot \hat{\bm{k}})^2 \nonumber\\
    &\hspace{1em}
    -4(1+e)(\bm{C}\cdot \bm{c}_{12})\left(\bm{C}\cdot \bm{c}_{12}\right)(\bm{c}_{12}\cdot \hat{\bm{k}}),\label{eq:Delta_4}
\end{align}
\end{subequations}
respectively.
Substituting Eq.\ (\ref{eq:Delta_2}) into Eq.\ (\ref{eq:mu_n_}), we can calculate the second moment $\mu_2$ as
\begin{widetext}
\begin{align}
    \mu_2 &= \frac{1}{4}(1-e^2)\int \mathrm{d}\bm{c}_1 \int \mathrm{d}\bm{c}_2 \int \mathrm{d}\hat{\bm{k}}
    \tilde{\sigma}(\chi,c_{12}) c_{12}\phi(c_1)\phi(c_2)(\bm{c}_{12}\cdot \hat{\bm{k}})^2\nonumber\\
    &\hspace{1em}\times \left\{1+a_2\left[C^4 + (\bm{C}\cdot \bm{c}_{12})^2+\frac{1}{16}c_{12}^4
    +\frac{1}{2}C^2c_{12}^2 -5C^2-\frac{5}{4}c_{12}^2+\frac{15}{4}\right]\right\}\nonumber\\
    &=\frac{1}{4}(1-e^2)\left[J_{0,0,0,0,2}+a_2\left(J_{4,0,0,0,2}+J_{0,0,2,0,2}+\frac{1}{16}J_{0,4,0,0,2}+\frac{1}{2}J_{2,2,0,0,2}-5J_{2,0,0,0,2}-\frac{5}{4}J_{0,2,0,0,2}+\frac{15}{4}J_{0,0,0,0,2}\right)\right]\nonumber\\
    &\equiv \sqrt{2\pi} \left(\frac{\varepsilon}{T}\right)^{2/\alpha}\Gamma\left(3-\frac{2}{\alpha}\right)(S_1+a_2S_2),
\end{align}
where $S_1$ and $S_2$ are given by
\begin{subequations}\label{eq:S1_S2}
\begin{align}
    S_1 &= 2A_{\alpha,2}(1-e^2),\label{eq:S1}\\
    S_2 &=\frac{3\alpha^2-16\alpha+16}{8\alpha^2}A_{\alpha,2} (1-e^2),\label{eq:S2}
\end{align}
\end{subequations}
respectively, with $A_{\alpha,2} \equiv \int_0^\infty \mathrm{d}\tilde{b}\hspace{0.2em} \tilde{b} \cos^2\theta_\alpha$.

Similarly, using Eq.\ (\ref{eq:Delta_4}), we can obtain the fourth moment $\mu_4$ as
\begin{align}
    \mu_4 &= -\frac{1}{2}\int \mathrm{d}\bm{c}_1 \int \mathrm{d}\bm{c}_2 \int \mathrm{d}\hat{\bm{k}}
    \tilde{\sigma}(\chi,c_{12}) c_{12}\phi(c_1)\phi(c_2)\nonumber\\
    &\hspace{1em}\times \left\{1+a_2\left[C^4 + (\bm{C}\cdot \bm{c}_{12})^2+\frac{1}{16}c_{12}^4
    +\frac{1}{2}C^2c_{12}^2 -5C^2-\frac{5}{4}c_{12}^2+\frac{15}{4}\right]\right\}\nonumber\\
    &\hspace{1em}\times \left\{2(1+e)^2(\bm{C}\cdot \hat{\bm{k}})^2(\bm{c}_{12}\cdot \hat{\bm{k}})^2
    +\frac{1}{8}(1-e^2)^2 (\bm{c}_{12}\cdot \hat{\bm{k}})^4 -\frac{1}{4}(1-e^2)^2c_{12}^2 (\bm{c}_{12}\cdot \hat{\bm{k}})^2\right.\nonumber\\
    &\hspace{3em}\left.-(1-e^2)C^2(\bm{c}_{12}\cdot \hat{\bm{k}})^2 
    -4(1+e)(\bm{C}\cdot \bm{c}_{12})\left(\bm{C}\cdot \bm{c}_{12}\right)(\bm{c}_{12}\cdot \hat{\bm{k}})\right\}\nonumber\\
    &=-(1+e)^2 \left[J_{0,0,0,2,2} + a_2\left(J_{4,0,0,2,2}+J_{0,0,2,2,2}+\frac{1}{16}J_{0,4,0,2,2}+\frac{1}{2}J_{2,2,0,2,2}
    -5J_{2,0,0,2,2}-\frac{5}{4}J_{0,2,0,2,2}+\frac{15}{4}J_{0,0,0,2,2}\right)\right]\nonumber\\
    &\hspace{1em}-\frac{1}{16}(1-e^2)^2
    \left[J_{0,0,0,0,4}+a_2\left(J_{4,0,0,0,4}+J_{0,0,2,0,4}+\frac{1}{16}J_{0,4,0,0,4}+\frac{1}{2}J_{2,2,0,0,4}
    -5J_{2,0,0,0,4}-\frac{5}{4}J_{0,2,0,0,4}+\frac{15}{4}J_{0,0,0,0,4}\right)\right]\nonumber\\
    &\hspace{1em}+\frac{1}{8}(1-e^2)^2\left[J_{0,2,0,0,2} + a_2\left(J_{4,2,0,0,2}+J_{0,2,2,0,2}+\frac{1}{16}J_{0,6,0,0,2}+\frac{1}{2}J_{2,4,0,0,2}
    -5J_{2,2,0,0,2}-\frac{5}{4}J_{0,4,0,0,2}+\frac{15}{4}J_{0,2,0,0,2}\right)\right]\nonumber\\
    &\hspace{1em}+\frac{1}{2}(1-e^2) \left[J_{2,0,0,0,2} + a_2\left(J_{6,0,0,0,2}+J_{2,0,2,0,2}+\frac{1}{16}J_{2,4,0,0,2}+\frac{1}{2}J_{4,2,0,0,2}
    -5J_{4,0,0,0,2}-\frac{5}{4}J_{2,2,0,0,2}+\frac{15}{4}J_{2,0,0,0,2}\right)\right]\nonumber\\
    &\hspace{1em}+2(1+e) \left[J_{0,0,1,1,1} + a_2\left(J_{4,0,1,1,1}+J_{0,0,3,1,1}+\frac{1}{16}J_{0,4,1,1,1}+\frac{1}{2}J_{2,2,1,1,1}
    -5J_{2,0,1,1,1}-\frac{5}{4}J_{0,2,1,1,1}+\frac{15}{4}J_{0,0,1,1,1}\right)\right]\nonumber\\
    &\equiv \sqrt{2\pi} \left(\frac{\varepsilon}{T}\right)^{2/\alpha}\Gamma\left(3-\frac{2}{\alpha}\right)(T_1+a_2T_2),
\end{align}
where $T_1$ and $T_2$ are given by
\begin{subequations}\label{eq:T1_T2}
\begin{align}
    T_1 &=\frac{1}{\alpha}\left[(11\alpha-4)A_{\alpha,2} -(3\alpha-2)A_{\alpha,4} (1-e^2)\right](1-e^2),\label{eq:T1}\\
    T_2 &=\frac{1}{16\alpha^3}\left[(173\alpha^3-444\alpha^2+304\alpha-64)A_{\alpha,2}
    + (51\alpha^3+62\alpha^2-112\alpha+32)A_{\alpha,4}\right.\nonumber\\
    &\hspace{4em}\left.+(45\alpha^3-126\alpha^2+112\alpha-32)A_{\alpha,4}e^2\right](1-e^2)\nonumber\\
    &\hspace{1em}+\frac{4}{\alpha}(3\alpha-2)(A_{\alpha,2}-A_{\alpha,4})(1+e),\label{eq:T2}
\end{align}
\end{subequations}
respectively, with $A_{\alpha,4} \equiv \int_0^\infty \mathrm{d}\tilde{b}\hspace{0.2em} \tilde{b} \cos^4\theta_\alpha$.

Next, let us obtain the explicit expressions of $\Omega_\eta^e$ and $\Omega_\kappa^e$ to derive the transport coefficients.
From the definition of $\tilde{D}_{ij}(\bm{c})$, we can rewrite $\tilde{D}_{ij}(\bm{c}_2)\Delta\left[\tilde{D}_{ij}(\bm{c}_1)+\tilde{D}_{ij}(\bm{c}_2)\right]$ as
\begin{align}
    &\tilde{D}_{ij}(\bm{c}_2)\Delta [\tilde{D}_{ij}(\bm{c}_1)+\tilde{D}_{ij}(\bm{c}_2)]\nonumber\\
    &=\left(c_{2i}c_{2j}-\frac{1}{3}\delta_{ij}c_2^2\right) \left[ c_{1i}^\prime c_{1j}^\prime + c_{2i}^\prime c_{2j}^\prime 
    -c_{1i} c_{1j} - c_{2i} c_{2j} -\frac{1}{3}\delta_{ij}\left(c_1^{\prime2}+c_2^{\prime2}-c_1^2-c_2^2\right)\right]\nonumber\\
    &=\frac{1-e^2}{6}C^2(\bm{c}_{12}\cdot \hat{\bm{k}})^2 + \frac{1+e}{2}c_{12}^2 (\bm{C}\cdot \hat{\bm{k}})(\bm{c}_{12}\cdot \hat{\bm{k}})
    -\frac{(1+e)(5+e)}{24} c_{12}^2 (\bm{c}_{12}\cdot \hat{\bm{k}})^2- (1+e)(\bm{C}\cdot \bm{c}_{12})(\bm{C}\cdot \hat{\bm{k}})(\bm{c}_{12}\cdot \hat{\bm{k}}) \nonumber\\
    &\hspace{1em}+\frac{(1+e)(2+e)}{6}(\bm{C}\cdot \bm{c}_{12})(\bm{c}_{12}\cdot \hat{\bm{k}})^2+\frac{(1+e)^2}{2}(\bm{C}\cdot \hat{\bm{k}})^2(\bm{c}_{12}\cdot \hat{\bm{k}})^2
    -\frac{(1+e)^2}{2} (\bm{C}\cdot \hat{\bm{k}})(\bm{c}_{12}\cdot \hat{\bm{k}})^3
    +\frac{(1+e)^2}{8}(\bm{c}_{12}\cdot \hat{\bm{k}})^4.\label{eq:D_delta_D}
\end{align}
Substituting Eq.\ (\ref{eq:D_delta_D}) into the definition of $\Omega_\eta^e$, we can obtain
\begin{align}
    \Omega_\eta^e&=-\sqrt{2\pi}\left(\frac{\varepsilon}{T}\right)^{2/\alpha}\Gamma\left(4-\frac{2}{\alpha}\right)
    \left(\omega_{\eta,1}+a_2\omega_{\eta,2}\right),
\end{align}
with
\begin{subequations}\label{eq:omega_eta1_eta2}
\begin{align}
    \omega_{\eta,1}&\equiv\frac{2}{3}(1+e)
    \left[5A_{\alpha,2}-3A_{\alpha,4} + (A_{\alpha,2}-3A_{\alpha,4})e\right],\label{eq:omega_eta1}\\
    \omega_{\eta,2}&\equiv-\frac{2(\alpha^2-16)}{96\alpha^2}(1+e)\left[5A_{\alpha,2}-3A_{\alpha,4} + (A_{\alpha,2}-3A_{\alpha,4})e\right].\label{eq:omega_eta2}
\end{align}
\end{subequations}

Similarly, from the definition of $\tilde{\bm{S}}(\bm{c})$, we can rewrite $\tilde{\bm{S}}(\bm{c}_2)\cdot \Delta [\tilde{\bm{S}}(\bm{c}_1)+\tilde{\bm{S}}(\bm{c}_2)]$ as
\begin{align}
    &\tilde{\bm{S}}(\bm{c}_2)\cdot \Delta \left[\tilde{\bm{S}}(\bm{c}_1)+\tilde{\bm{S}}(\bm{c}_2)\right]\nonumber\\
    &=\left(c_2^2-\frac{5}{2}\right)\left[ (\bm{c}_1^\prime \cdot \bm{c}_2)c_1^{\prime2}+(\bm{c}_2^\prime \cdot \bm{c}_2)c_2^{\prime2}
    -(\bm{c}_1 \cdot \bm{c}_2)c_1^2-(\bm{c}_2 \cdot \bm{c}_2)c_2^2 \right]\nonumber\\
    &=-\frac{1-e^2}{2}C^4(\bm{c}_{12}\cdot \hat{\bm{k}})^2 +\frac{1+e}{2}C^2c_{12}^2(\bm{C}\cdot \hat{\bm{k}})(\bm{c}_{12}\cdot \hat{\bm{k}})
    -\frac{1-e^2}{8}C^2c_{12}^2(\bm{c}_{12}\cdot \hat{\bm{k}})^2-2(1+e)C^2(\bm{C}\cdot\bm{c}_{12})(\bm{C}\cdot \hat{\bm{k}})(\bm{c}_{12}\cdot \hat{\bm{k}}) \nonumber\\
    &\hspace{1em}
    +\frac{(1+e)(5-3e)}{4}C^2(\bm{C}_{12}\cdot \bm{c}_{12})(\bm{c}_{12}\cdot \hat{\bm{k}})^2
    +(1+e)^2C^2(\bm{C}\cdot \hat{\bm{k}})^2(\bm{c}_{12}\cdot \hat{\bm{k}})^2
    -\frac{(1+e)^2}{2}C^2(\bm{C}\cdot \hat{\bm{k}}) (\bm{c}_{12}\cdot \hat{\bm{k}})^3\nonumber\\
    &\hspace{1em}+\frac{5(1-e^2)}{4}C^2(\bm{c}_{12}\cdot \hat{\bm{k}})^2
    +\frac{1+e}{8}c_{12}^4(\bm{C}\cdot \hat{\bm{k}}) (\bm{c}_{12}\cdot \hat{\bm{k}})
    -(1+e)c_{12}^2(\bm{C}\cdot \bm{c}_{12})(\bm{C}\cdot \hat{\bm{k}}) (\bm{c}_{12}\cdot \hat{\bm{k}})\nonumber\\
    &\hspace{1em}+\frac{(1+e)(3-e)}{16} c_{12}^2(\bm{C}\cdot \bm{c}_{12})(\bm{c}_{12}\cdot \hat{\bm{k}})^2
    +\frac{(1+e)^2}{4}c_{12}^2 (\bm{C}\cdot \hat{\bm{k}})^2(\bm{c}_{12}\cdot \hat{\bm{k}})^2
    -\frac{(1+e)^2}{8}c_{12}^2 (\bm{C}\cdot \hat{\bm{k}}) (\bm{c}_{12}\cdot \hat{\bm{k}})^3\nonumber\\
    &\hspace{1em}
    -\frac{5(1+e)}{4}c_{12}^2(\bm{C}\cdot \hat{\bm{k}}) (\bm{c}_{12}\cdot \hat{\bm{k}})
    +2(1+e)(\bm{C}\cdot \bm{c}_{12})^2 (\bm{C}\cdot \hat{\bm{k}})(\bm{c}_{12}\cdot \hat{\bm{k}})
    -\frac{(1+e)(3-e)}{4}(\bm{C}\cdot \bm{c}_{12})^2(\bm{c}_{12}\cdot \hat{\bm{k}})^2\nonumber\\
    &\hspace{1em}-(1+e)^2(\bm{C}\cdot \bm{c}_{12}) (\bm{C}\cdot \hat{\bm{k}})^2(\bm{c}_{12}\cdot \hat{\bm{k}})^2
    +\frac{(1+e)^2}{2}(\bm{C}\cdot \bm{c}_{12}) (\bm{C}\cdot \hat{\bm{k}})(\bm{c}_{12}\cdot \hat{\bm{k}})^3
    +5(1+e)(\bm{C}\cdot \bm{c}_{12}) (\bm{C}\cdot \hat{\bm{k}})(\bm{c}_{12}\cdot \hat{\bm{k}})\nonumber\\
    &\hspace{1em}-\frac{5(1+e)(3-e)}{8}(\bm{C}\cdot \bm{c}_{12}) (\bm{c}_{12}\cdot \hat{\bm{k}})^2
    -\frac{5(1+e)^2}{2} (\bm{C}\cdot \hat{\bm{k}})^2 (\bm{c}_{12}\cdot \hat{\bm{k}})^2
    +\frac{5(1+e)^2}{4} (\bm{C}\cdot \hat{\bm{k}})(\bm{c}_{12}\cdot \hat{\bm{k}})^3.\label{eq:S_delta_S}
\end{align}
\end{widetext}
Substituting Eq.\ (\ref{eq:S_delta_S}) into the definition of $\Omega_\kappa^e$, we can obtain
\begin{align}
    \Omega_\kappa^e
    &=-\sqrt{2\pi}\left(\frac{\varepsilon}{T}\right)^{2/\alpha}\Gamma
    \left(4-\frac{2}{\alpha}\right)\left(\omega_{\kappa,1}+a_2\omega_{\kappa,2}\right),
\end{align}
with
\begin{subequations}\label{eq:omega_kappa1_kappa2}
\begin{align}
    \omega_{\kappa,1}
    &\equiv \frac{1+e}{4(3\alpha-2)}
    \left\{5(13\alpha-12)A_{\alpha,2}-8(3\alpha-2)A_{\alpha,4}\right.\nonumber\\
    &\hspace{6em}\left.-[(17\alpha-28)A_{\alpha,2}+8(3\alpha-2)A_{\alpha,4}]e \right\},\label{eq:omega_kappa1}\\
    \omega_{\kappa,2}
    &\equiv \frac{\alpha^2-16}{128\alpha^2(3\alpha-2)}(1+e)\nonumber\\
    &\hspace{1em}\times \left\{5(7\alpha-4)A_{\alpha,2}-8(3\alpha-2)A_{\alpha,4}\right.\nonumber\\
    &\hspace{2.5em}\left.
    +[(13\alpha-12)A_{\alpha,2}-8(3\alpha-2)A_{\alpha,4}]e\right\}.\label{eq:omega_kappa2}
\end{align}
\end{subequations}


\end{document}